\documentclass{article}
\usepackage{smc2021}
\usepackage[caption=false, font=footnotesize]{subfig}
\usepackage{paralist}
\usepackage[figure,table]{hypcap}

\usepackage[whole]{bxcjkjatype}

\usepackage{alphabeta}
\usepackage{arabtex}
\usepackage[LFE,LAE,LGR,T2A,T1]{fontenc}
\usepackage[greek, russian, main=english]{babel}

\def\papertitle{Musical Prosody-Driven Emotion Classification: Interpreting Vocalists Portrayal of Emotions Through Machine Learning}
\author[1]{\mbox{\firstname{Nicholas}\lastname{Farris}\email{nicholas.farris@gatech.edu}}}
\author[1]{\mbox{\firstname{Brian}\lastname{Model}\email{bmodel@gatech.edu}}}
\author[1]{\mbox{\firstname{Richard}\lastname{Savery}\email{rsavery3@gatech.edu}}}
\author[1]{\mbox{\firstname{Gil}\lastname{Weinberg}\email{gilw@gatech.edu}}}

\affil[1]{\department{Robotic Musicianship Lab}\institution{Georgia Institute of Technology} \city{Atlanta}\state{GA}\country{USA}\affiliationtype{University}}

\completesetup

\title{\papertitle}
\begin{document}
	\capstartfalse
	\maketitle
	\capstarttrue

	\begin{abstract}
	
	The task of classifying emotions within a musical track has received widespread attention within the Music Information Retrieval (MIR) community. Music emotion recognition has traditionally relied on the use of acoustic features, verbal features, and metadata-based filtering. The role of musical prosody remains under-explored despite  several studies demonstrating a strong connection between prosody and emotion. In this study, we restrict the input of traditional machine learning algorithms to the features of musical prosody. Furthermore, our proposed approach builds upon the prior by classifying emotions under an expanded emotional taxonomy, using the Geneva Wheel of Emotion. We utilize a methodology for individual data collection from vocalists, and personal ground truth labeling by the artist themselves. We found that traditional machine learning algorithms when limited to the features of musical prosody (1) achieve high accuracies for a single singer, (2) maintain high accuracy when the dataset is expanded to multiple singers, and (3) achieve high accuracies when trained on a reduced subset of the total features.
	
	\end{abstract}

	\section{Introduction}\label{sec:introduction}
	
    The work presented in this paper is situated in the intersection between research on emotion for robotics \cite{savery2020} and emotional classification research in Music Information Retrieval \cite{cano2005content}. In particular, we focus on the under-explored domain of emotion-driven prosody for human-robot interaction \cite{savery2019establishing}. Verbal prosody is concerned with elements of speech that are not individual phonetic segments but rather pertain to linguistic functions such as intonation, tone, stress, and rhythm. Similarly, musical prosody is defined as the performer's manipulation of music for certain expressive and coordinating functions \cite{palmer06}. It has been hypothesized that these expressive functions serve to communicate emotion \cite{juslin2001music}. 
    
	In this paper, we explore the relationship between musical prosody and emotion through three research questions. First, are traditional machine learning algorithms able to accurately classify an individual's emotions when trained on only the features of musical prosody? Next, are these models able to generalize to a larger group of vocalists? Finally, which features of musical prosody contribute the most to the classification of emotion?
	
	The paper is structured as follows, in Section \ref{sec:background_and_motivation}, background and motivation are discussed. Section \ref{sec:metholdology} describes the dataset collection, training and testing, the taxonomies used in classification, the feature extraction methodology and analysis of their relevance to emotion, feature aggregation, feature selection, and model generalization. Section \ref{sec:results} presents the experiments: Experiment 1 asks how well can traditional machine learning models classify emotion when limited to inputs of musical prosody, Experiment 2 explores our approach's ability to generalize to a larger population of singers, and Experiment 3 explores the individual contribution to accuracy of each feature via training on reduced subsets of the input vector. Section \ref{sec:discussion} provides discussion to these results, with particular attention paid to the relationships between emotions and potential future work. Finally, section \ref{sec:conclusions} concludes the paper. A demo via python notebook with audio samples is available online. \footnote{\url{https://github.com/brianmodel/EmotionClassification}}
	
	\section{Background}\label{sec:background_and_motivation}
	
	Emotion classification has been a major focus of research in recent years. Ekman created a discrete categorization that consists of fundamental basic emotions which are the root for more complex emotions \cite{ekman}. Another classification model is the Circumplex model proposed by Posner et al which plots emotions on a continuous, two-dimensional scale of valence and arousal \cite{posner}. In this paper, we classify emotions using a model similar to the two-dimensional Circumplex model which is further described in section 3.1.
	
    There has also been much work done in the field of analyzing emotion from text for tasks such as sentiment analysis. Research on classification of emotion in audio has taken many different approaches. Research into classifying emotions in knocking sounds has found that anger, happiness and sadness could be easily classified from audio alone \cite{Houel1446843}. There have been multimodal approaches which use audio in combination with another feature, namely visual facial features \cite{haq2008}\cite{840655} or text lyrics \cite{DBLP:journals/corr/JamdarAKD15}. Furthermore, researchers have performed emotional classification from audio in the context of music by analyzing which musical features best convey emotions \cite{song2012}. Panda et al. have found a relationship between melodic and  dynamic features to  a number of specific emotions \cite{9229494}. Such features that were used to classify emotion in music, however, cannot be easily generalized to other domains. Prosody has been found by linguists to communicate emotion across various cultures, with patterns of pitch and loudness over time representing different emotions \cite{frick1985}, and has shown the potential to improve human-robot interaction \cite{savery2019finding,savery2020emotional,saverybetween}. Our approach aims to bridge this gap by analyzing these prosodic features which are fundamental to everyday speech and explore how they can be used to classify emotional driven prosody. 
    
    Koo et al. have done work in speech emotion recognition using a combination of MFCC and prosodic features with a GRU model on the IEMOCAP dataset \cite{9051281}. We expand upon their work by performing an in-depth analysis of 11 different audio features and their effect on classifying emotion. We also classify emotion beyond spoken language by analyzing prosodic features which better generalize to how humans convey emotion using the new dataset collected, as described in section 3.2.
	
	\section{Methodology}\label{sec:metholdology}
	
	\subsection{Taxonomy}
	
	One of the main challenges in emotional classification is the derivation of a taxonomy that accurately reflects the problem domain. The two common approaches to address this challenge are 1. Discrete emotional categorization; and 2. Continuous quantitative metrics of Valence and Arousal (sometimes called Control). We use both approaches with a categorical, as opposed to regression, approach to the latter. 
	
\begin{figure}[htp]
        \centering
        \includegraphics[width=8cm]{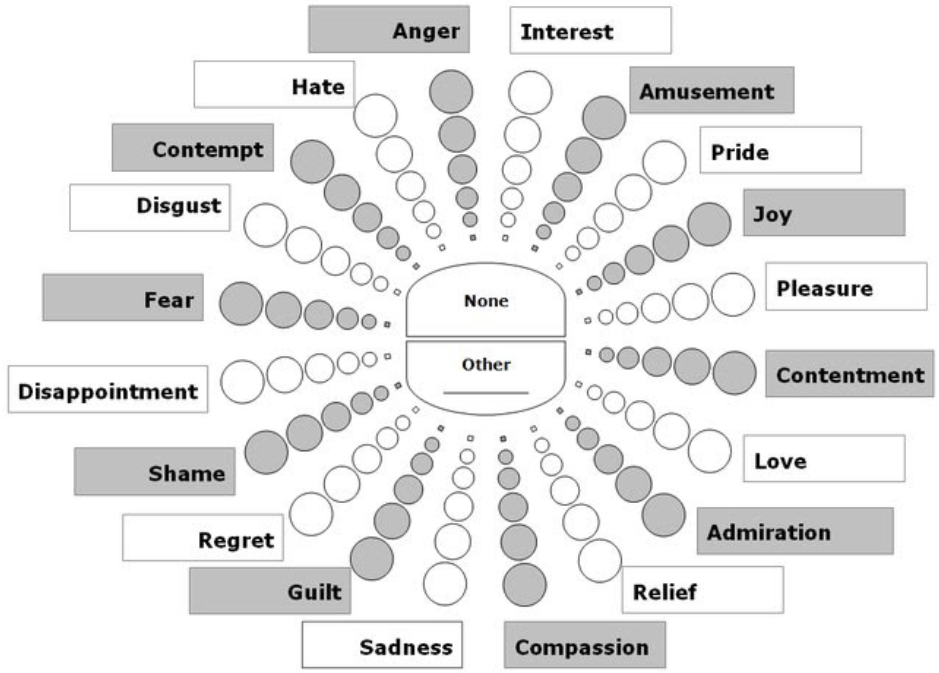}
        \caption{The Geneva Wheel of Emotion}
        \label{fig:wheelOfEmotion}
    \end{figure}

	Our models classify emotion under two taxonomies: first we categorize each data point as belonging to one of the twenty emotions located around the Geneva Wheel of Emotion. Then we categorize each data point as belonging to one of the quadrants depicted by the intersection of valance and control by assigning each emotion from the Geneva Wheel of Emotion to its respective quadrant. We abbreviate each of these quadrants as follows: {"High Control Negative Valance": "HCN"}, {"High Control Positive Valance": "HCP"}, {"Low Control Negative Valance": "LCN"}, and {"Low Control Positive Valance": "LCP"}. See Table \ref{table:taxonomy} and Figure \ref{fig:wheelOfEmotion} for a visualization of the domain's taxonomy.

    \begin{table}[h!]
    \centering
    	\begin{tabular}{ |p{1.3cm}|p{1.6cm}|p{2.1cm}|p{1.7cm}|  }
            \hline
            HCN&HCP&LCN&LCP\\
            \hline
            Anger&Amusement&Disappointment&Admiration\\
            Contempt&Interest&Fear&Compassion\\
            Disgust&Joy&Guilt&Contentment\\
            Hate&Pleasure&Sadness&Love\\
            Regret&Pride&Shame&Relief\\
            \hline
        \end{tabular}
    \caption{Selected emotional taxonomy for training}
    \label{table:taxonomy}
    \end{table}
    
	\subsection{Data Collection}

    Due to a lack of data labeled with the appropriate taxonomy, we decided to collect and annotate a new dataset. To achieve this goal, we asked professional singers to consciously sing each emotion. To generate our dataset, three professional singers were tasked to improvise as many phrases as possible for each emotion in the Geneva Wheel of Emotion. The singers were instructed to sing each phrase between 1 and 20 seconds, and to spend approximately 15 minutes on each emotion, resulting in 4 to 6 hours of recordings per singer annotated with ground-truth labels.
    
    Additionally, the singers were given the following instructions during their recording session:
    \begin{enumerate}
        \item Do not attempt to control for different intensities for each emotion
        \item  Sing anything for each phrase that you believe matches the emotion except use words.
        \item After recording, mark any phrase that you believe did not capture the intended emotion and it will be deleted
   
    \end{enumerate}
	
	\subsection{Feature Extraction}
	
	In the following section, we define the features selected for extraction from our dataset prior to model training. Furthermore, we discuss each feature's relevance to emotional classification through an analysis of prior works.
    
    \subsubsection{Zero Crossing Rate}
    
    Zero Crossing Rate, the rate of sign-changes across a signal, is key in classifying percussive sounds. Unvoiced regions of audio are known to have higher Zero Crossing Rates \cite{853637}. One study analyzed ZCR for Anger, Fear, Neutral, and Happy signals and noted that higher peaks were found for Happy and Anger emotions \cite{8228105}.
	
    \subsubsection{Energy}
    
    Energy, the area under the squared magnitude of the considered signal, relates to the amount of spectral information in a signal \cite{peeters2004large} and previous studies have found energy is essential in distinguishing stressed and neutral speech \cite{kwon2003emotion}.
    
    \subsubsection{Entropy of Energy}
    
    Entropy of Energy, the average level of "information" or "uncertainty" inherent within a signal's energy, has been shown in one study to have similar values for disgust and boredom \cite{lalitha2015speech}. To accurately measure the entropy of the different emotions, we must make sure we are not including parts of the signal where the individual is not speaking.
    
    \subsubsection{Spectral Centroid}
    
    Spectral Centroid, the power spectrum's center of mass, perceptually has a connection with a sound's brightness. It follows, that this parameter serves as an indicator of musical timbre \cite{kendall1996difference}. Previous studies have shown spectral centroid is a significant component in music emotion \cite{wu2014musical}.
    
    \subsubsection{Spectral Spread}
    
    Spectral Spread, the second central moment of the power spectrum, has shown to help the listener to differentiate noise-like and tone-like portions of a signal \cite{jain2018cubic}.
    
    \subsubsection{Spectral Entropy}
    
    Spectral Entropy, the entropy of the power spectrum, when used with MFCC features has shown an improvement in speech recognition accuracy \cite{toh2005}. Another study found spectral entropy to have the highest correlation to emotional valence of all features tested \cite{10.1145/3243274.3243313}.
    
    \subsubsection{Spectral Flux}
    
    Spectral Flux, a measure of the rate of change of the power spectrum calculated as the Euclidean distance between sequential frames, relates to how fast the pitch changes in time and has been shown to be dominant in cross-domain emotion recognition from speech and sound and from sound and music \cite{weninger2013acoustics}.
    
    \subsubsection{Spectral Rolloff}
    
    Spectral Rolloff, the frequency under which some percentage of the total energy of the spectrum is contained, helps differentiate between harmonic content, characterized below the roll-off, and noisy sounds, characterized above the roll-off. Spectral rolloff has been shown to be one of the most important prosodic features in classifying emotion \cite{10.1145/3243274.3243313}. 
    
    \subsubsection{MFCCs}
    
    Mel-Frequency Cepstral Coefficients (MFCCs), a representation of the short-term power spectrum based on a linear cosine transform of a log power spectrum on a nonlinear mel scale of frequency, are used in speech recognition with their ability to represent the speech amplitude spectrum in a compact form \cite{logan2000mel}. Many studies have linked the importance of MFCC analysis to emotion recognition \cite{8228105} \cite{6514336} \cite{neiberg2006emotion} .
    
    \subsubsection{Chroma Vector and Deviation}
    
    Chroma Vector, an approximation of the pitch class profiles present within a given frame and often used as the twelve tones, allows for the capture of harmonic and melodic characteristics while remaining robust toward changes in timbre and instrumentation. Previous studies have shown increases in emotional classification accuracy with chroma vector and its standard deviation \cite{kim2010,schmidt2011learning}.
	
	\subsection{Feature Aggregation}
	
    \begin{figure}[htp]
        \centering
        \includegraphics[width=8cm]{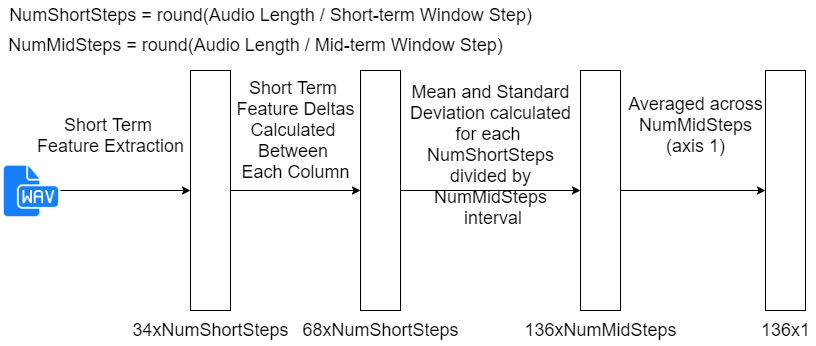}
        \caption{Model of Feature Aggregation}
        \label{fig:featureAggregation}
    \end{figure}
    
	\begin{table}
	\centering
        \begin{tabular}{ |c|c| } 
            \hline
            Parameter & Value \\
            \hline
            Mid-term Window Step & 1.0 seconds \\ 
            Mid-term Window Size & 1.0 seconds \\ 
            Short-term Window Step & 0.05 seconds \\ 
            Short-term Window Size & 0.05 seconds \\ 
            \hline
        \end{tabular}
        \caption{Feature Aggregation Parameters}
        \label{table:aggregationParameters}
    \end{table}
    
    In this section, we define the aggregation pipeline from feature extraction to feature vector for each audio file. Figure \ref{fig:featureAggregation} provides a visual modeling of our feature aggregation pipeline. Table \ref{table:aggregationParameters} delineates the feature aggregation hyper parameters used in this study.
    
    \subsubsection{Short-term Aggregation}
    
    The short-term aggregation of a 5-second clip, using a Short-term Window Step of .05 seconds and a Short-term Window Size of .05 seconds is defined as follows: Each of the 34 features discussed above are extracted for every 50ms, resulting in 100 feature vectors of size 34x1, represented as a 34x100 matrix. Next, the deltas between each time step are calculated according to the equation $delta = feature\_vector - feature\_vector\_prev$. The first time stamp has all deltas set to 0. Each delta vector is concatenated onto its respective feature vector resulting in a size of 68x1, represented as a 68x100 matrix for the entire 5 second audio clip.
    
    \subsubsection{Mid-term Aggregation}
    
    Next, mid-term aggregation occurs with a Mid-term Window Size of 1.0 seconds and Mid-term Window Step of 1.0 seconds. The 68x100 matrix of Short-term features is split according to the ratio between the Mid-term and Short-term window size and step, resulting in 5 matrices of size 68x20. For each matrix, we calculate and flatten the mean and standard deviation for each row, resulting in 5 136x1 mid-term feature vectors, represented as a 136x5 matrix. Finally, we take the mean across the first axis resulting in a 136x1 feature vector representing our 5 second audio clip.

	\subsection{Classification}

    Prior work focused on musical classification has primarily found success in the implementation of k-nearest neighbor (K-NN) and support vector machines (SVM), finding the highest accuracies using SVMs \cite{bischoff2009}. In exploration of the relationship between musical prosody and emotion, we will implement a variety of machine learning models, namely we will train and evaluate KNNs, Linear SVMs, Random Forests, Extra Trees, Gradient Boosting, and Feed Forward Neural Networks (FFNN). FFNNs are used in experiment 3 only.
    
    \textit{Experiment 1:} we explore the base line accuracies, F-scores, and confusion matrices achieved by training each model with identical training, validation, and testing data from a single singer.
    
    \textit{Experiment 2:} we explore our model architecture's ability to generalize by expanding the dataset to include all 3 singers from data collection.

    \textit{Experiment 3:} we explore model performance on a reduced subset of the training feature, utilizing additive feature selection to compile a ranking of features.
    
	\section{Results}\label{sec:results}
	
	\subsection{Experiment 1}
	
    \begin{figure*}[htp]
        \centering
        \includegraphics[width=17.2cm]{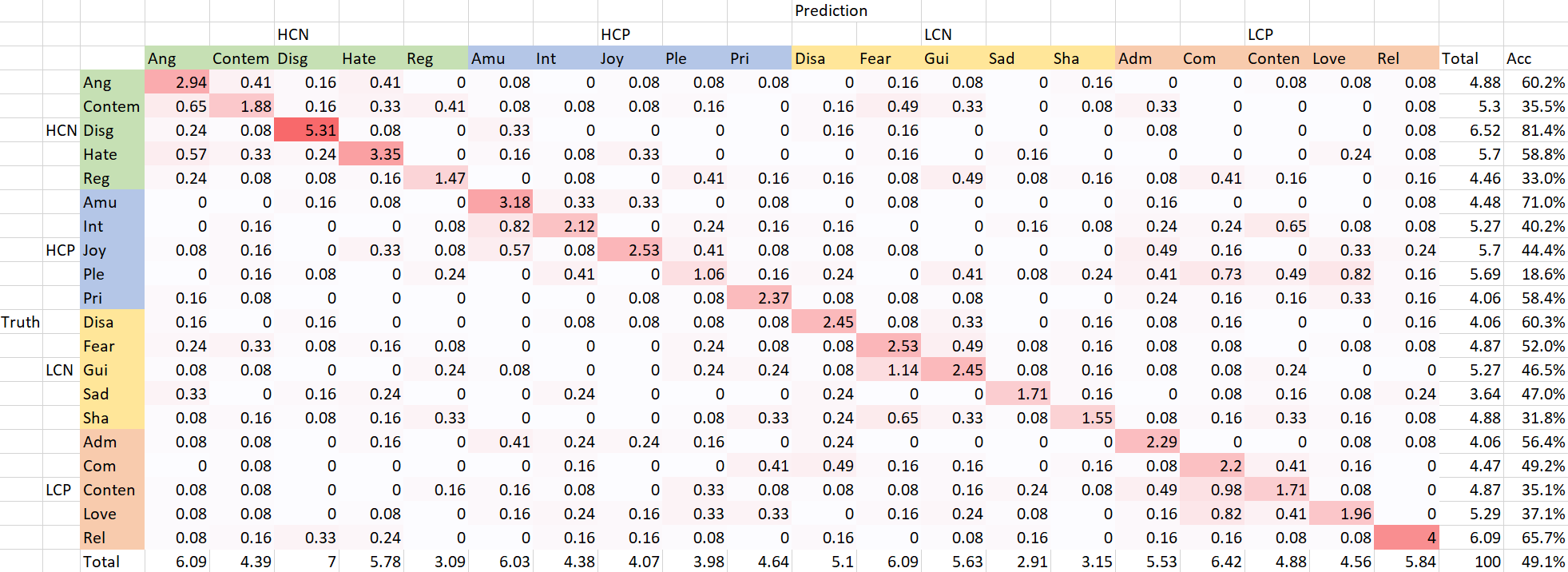}
        \caption{SVM, Individual Taxonomy, 1 Singers Confusion Matrix}
        \label{fig:individualConfusion}
    \end{figure*}
    
    \begin{figure}[h!]
        \centering
        \includegraphics[width=8cm]{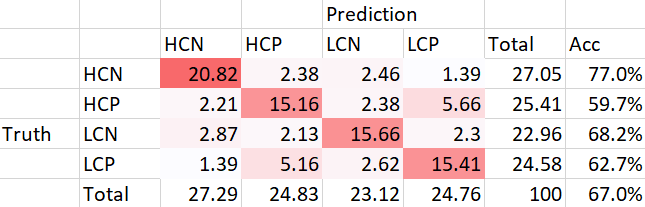}
        \caption{Gradient Boosting, Big 4 Taxonomy, 1 Singer Confusion Matrix}
        \label{fig:GB_Big4_1S_Confusion}
    \end{figure}

	In experiment 1, we analyze the baseline accuracies, F-scores, and confusion matrices achieve by training KNNs, linear SVMs, Random Forests, Extra Trees, Gradient Boosting models on a single singer utilizing only the prosodic features outlined in the previous section. All models were trained with features extracted according to the parameters outlined in Table \ref{table:aggregationParameters}. Additionally, each model is optimized with respect to its associated hyper parameter. We optimize KNN for the number nearest neighbors, SVM for the soft margin, random forest for number of trees, gradient boosting for the number of boosting stages, and extra trees for the number of trees. \footnote{\url{https://scikit-learn.org/}}
	
	Table \ref{table:Big4SingleSinger} provides the best accuracy, F1-score, and selected hyper-parameter for each of our models trained on a Big 4 taxonomy for a single singer. All models perform better than twice the accuracy of random guessing, with the linear SVM and Gradient Boosting models achieving the highest accuracies. Further analysis of the confusion matrix of the Gradient Boosting model, shown in Figure \ref{fig:GB_Big4_1S_Confusion}, provides information about the classes that are most often confused for one another. The model struggles in distinguishing between Low Control Positive Valance and High Control Positive Valance. This is to say the model can tell that an individual is in a positive mood, but has difficulties distinguishing the Control or Arousal of the emotion. 
	
	Next, we examine classification under a single emotion taxonomy for a single singer. Table \ref{table:IndividualSingleSinger} shows the best accuracy, F1-score, and selected hyper-parameter for each of our models. Each model significantly outperforms random guessing. Even the worst model, the KNN, performs 6.5 times better than random chance (20 possible categories = $5\%$ chance random guessing). Our best model, the linear SVM, performs approximately 10 times better than random guessing with an accuracy of 49.1\%. The confusion matrix for the single emotion taxonomy has been included in Figure \ref{fig:individualConfusion}. Analysis of this confusion matrix yields a few  observations: Disgust is rarely confused with other emotions, having the highest individual accuracy of 81.4\%. Fear and Guilt are the two most common pair of emotions to be confused for one another. Pleasure is the most difficult emotion for the model to classify correctly, having the lowest individual accuracy of 18.6\%.
	
	Finally, our models perform extremely well when tasked with categorizing between two emotions, achieving accuracies as high as 98.9\% with a f1 of 98.9 in the distinction between Love and Disgust using a SVM. This reinforces the intuition that by reducing the number of emotional categories we can achieve higher accuracies for identification. 
	
    \begin{figure*}[h!]
        \centering
        \includegraphics[width=17.2cm]{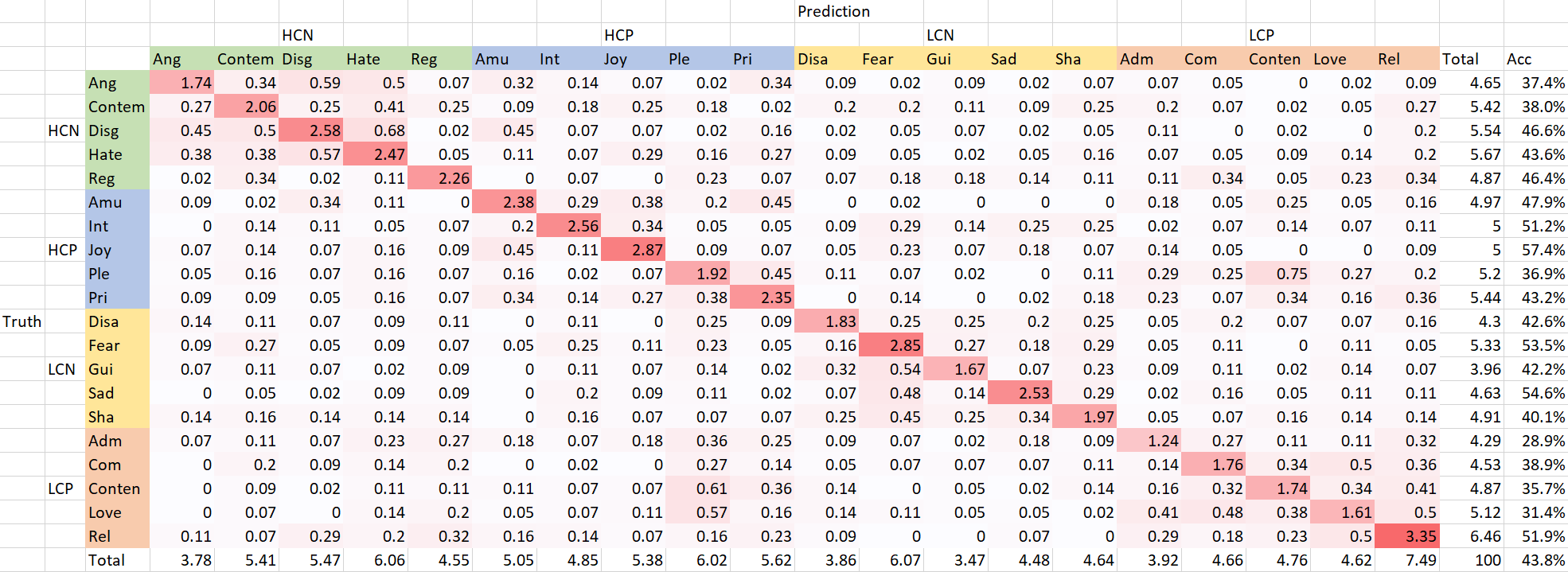}
        \caption{Gradient Boosting, Individual Taxonomy, 3 Singers Confusion Matrix}
        \label{fig:GB_Indiv_3S_Confusion}
    \end{figure*}
	
	\begin{table}
	\centering
        \begin{tabular}{ |c|c|c|c| } 
            \hline
            Model & Accuracy & F1 & Hyperparam \\
            \hline
            KNN & 56.1 & 56.2 & C=11 \\ 
            SVM & 66.5 & 65.3 & C=1.0 \\ 
            Extra Trees & 64.6 & 64.3 & C=100 \\ 
            Gradient Boosting & 67.0 & 66.7 & C=500 \\
            Random Forest & 63.5 & 63.2 & C=200 \\ 
            \hline
        \end{tabular}
        \caption{Big 4 Taxonomy, 1 Singer Classification Results}
        \label{table:Big4SingleSinger}
    \end{table}
    
	\begin{table}
	\centering
        \begin{tabular}{ |c|c|c|c| } 
            \hline
            Model & Accuracy & F1 & Hyperparam \\
            \hline
            KNN & 33.8 & 32.1 & C=15 \\ 
            SVM & 49.1 & 48.1 & C=5.0 \\ 
            Extra Trees & 44.3 & 42.8 & C=500 \\ 
            Gradient Boosting & 47.2 & 46.6 & C=200 \\
            Random Forest & 43.8 & 42.3 & C=200 \\ 
            \hline
        \end{tabular}
        \caption{Single Taxonomy, 1 Singer Classification Results}
        \label{table:IndividualSingleSinger}
    \end{table}

	\subsection{Experiment 2}
	
	Within machine learning, model generalization poses many challenges as models tend to memorize data and perform worse when exposed to new datasets. In experiment 2, we generalized our model by training on 3 different singers as opposed to training on one singer. Tables \ref{table:Big4Generalized} and \ref{table:SingleGeneralized} compare the accuracies achieved by the various model architectures for 3 singers vs 1 singer.
	
	With the exception of linear SVM, all model architectures maintain similar accuracies when trained on the 3 singer datasets. This maintenance of accuracy demonstrates the ability for traditional machine learning models to generalize well to a larger population when trained on only the features of musical prosody. We are unsure of why linear SVMs perform worse during generalization as compared to other models, seeing a drop of 6\% in Big 4 taxonomy and a drop of 13\% in single emotion taxonomy. This drop could potentially be a limitation in our methodology of only applying a linear kernel to SVM training, as perhaps an RBF or polynomial kernel would be better able to generalize to a larger population.
	
	The results of this experiment are encouraging to the development of a general model of emotional classification based on musical prosody as accuracy is maintained when the dataset is expanded to a larger portion of the overall population.
	
    \begin{figure}[htp]
        \centering
        \includegraphics[width=8cm]{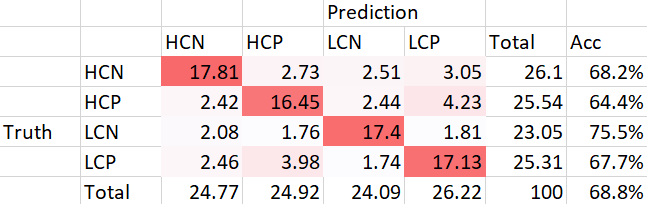}
        \caption{Gradient Boosting, Big 4 Taxonomy, 3 Singer Confusion Matrix}
        \label{fig:GB_Big4_3S_Confusion}
    \end{figure}
	
	\begin{table}
	\centering
        \begin{tabular}{ |c|c|c| } 
            \hline
            Model & 1-S Accuracy & 3-S Accuracy \\
            \hline
            KNN & 56.1 & 57.9 \\ 
            SVM & 66.5 & 60.6 \\ 
            Extra Trees & 64.6 & 63.5 \\ 
            Gradient Boosting & 67.0 & 68.8 \\
            Random Forest & 63.5 & 65.1 \\ 
            \hline
        \end{tabular}
        \caption{Big 4 Taxonomy, 1 Singer vs 3 Singer Accuracy}
        \label{table:Big4Generalized}
    \end{table}
    
	\begin{table}
	\centering
        \begin{tabular}{ |c|c|c| } 
            \hline
            Model & 1-S Accuracy & 3-S Accuracy \\
            \hline
            KNN & 33.8 & 32.5 \\ 
            SVM & 49.1 & 36.9 \\ 
            Extra Trees & 44.3 & 42.7 \\ 
            Gradient Boosting & 47.2 & 43.8 \\
            Random Forest & 43.8 & 43.8 \\ 
            \hline
        \end{tabular}
        \caption{Single Emotion Taxonomy, 1 Singer vs 3 Singer Accuracy}
        \label{table:SingleGeneralized}
    \end{table}
    
	\subsection{Experiment 3}

	Experiment 3 analyzes model performance on a reduced subset of the feature vector for our single emotion taxonomy. Our implementation of Feature Selection follows an additive approach. We start with an empty permanent feature set and each feature is trained on its own. The feature with the highest f1 score is selected and added to our permanent feature set. This process is repeated until all features have been added to the permanent feature set. Finally, we plot the f1 score vs features used in model training.
	
	For 136 features, an additive feature selection training loop requires the training and f1 validation of 9316 models. Our initial training and validation was based on implementations using the python library sklearn. Unfortunately, sklearn does not provide native GPU training support and thus performing an additive feature selection using sklearn is not feasible with respect to training time. Our solution is to continue to use the feature selection and aggregation outlined above, and to replace the sklearn models with a Tensorflow feed forward neural net. All of these models look for statistical correlations between our features and the emotional classification. Thus the particular model should have minimal affect on the analysis of feature importance performed by additive feature selection. Training was done sequentially on a RTX 3090 using CUDA v11 and took just under 24 hours to train and validate all 9316 models. 
	
	Our feed forward neural net contained the input layer, two dense layers of 136 nodes with relu activation functions, and a dense 20 node output layer. We trained using a Sparse Categorical Cross entropy loss function optimized using an Adam optimizer with 5 epochs per model.
	
\begin{figure*}[htp]
        \centering
        \includegraphics[width=17.2cm]{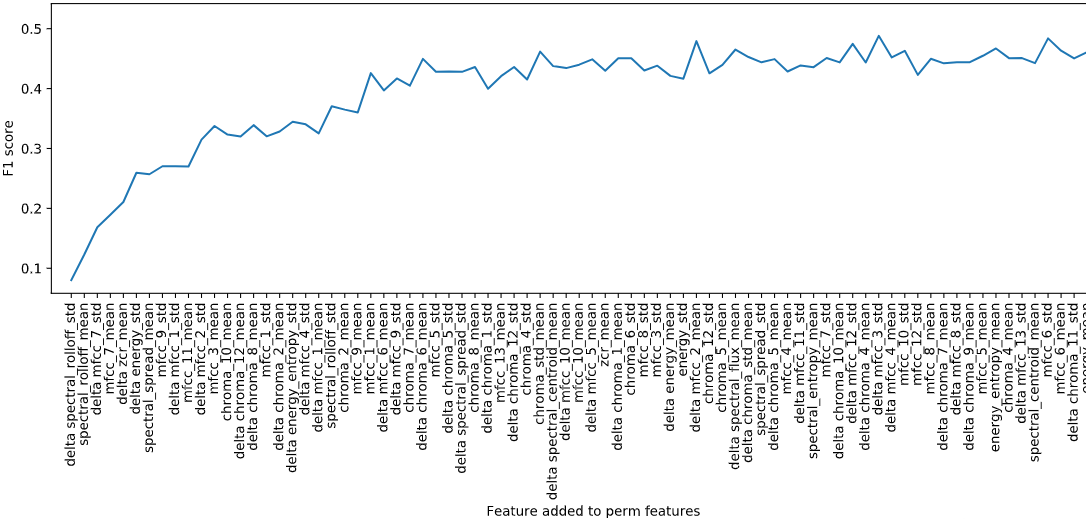}
        \caption{F1 score vs Features included in model pipeline}
        \label{fig:additiveFeatureSelection}
    \end{figure*}

	Figure \ref{fig:additiveFeatureSelection} shows the F1 score achieved vs the Feature included in the model pipeline. All feature on and to the right of any point in the x axis are included in training. An F1 of 45 is achieved within the first 25 features. Furthermore, the addition of the remaining 111 features only increases our F1 score to 52. This graph emphasizes the importance of spectral roll-off and MFCC 7 in the classification of emotion, as aggregations of these two features allow for an F1 score just below 20 with 4 total features.
	
	\section{Discussion}\label{sec:discussion}
	
	\subsection{Analysis}
    We demonstrate that prosodic features can be used to classify human emotions, achieving high accuracies on classifying emotions for a single singer dataset as seen in tables ~\ref{table:Big4SingleSinger} and ~\ref{table:IndividualSingleSinger}. Furthermore, we obtained encouraging results regarding  the model's generalization between singers as demonstrated by tables ~\ref{table:Big4Generalized} and ~\ref{table:SingleGeneralized}. However, given our limited dataset, more research is needed to study how the models generalize for additional singers with different voices.
    
    Our feature selection aligns with prior research indicating that energy and MFCC were the most useful features for classifying emotion \cite{haq2008}. However, we have been able to show that the results holds true not just for phonological speech, but in the more specific domain of musical prosody.
    
    \subsection{Relationships between Emotions}
    
	The classification results give us new insights into the uniqueness and relationships between emotions. Looking at the individual classification data between all the singers in Figure \ref{fig:individualConfusion}, we can see how the model was best able to classify fear, joy and relief. This is in contrast to emotions such pleasure or admiration which  showed the lowest classification accuracy. These results demonstrate the manner in which different humans convey emotions, and what  emotions are similarly expressed by different individuals. When conveying relief, all three singers expressed a diminuendo and exhale. Similarly, when conveying fear all three singers expressed a crescendo and more accented tones. On the other hand, there was a high level of variation when conveying pleasure, with many different tone ranges, mouth shapes, etc. being present in the data. 
	
	Furthermore, from the confusion matrix in Figure~\ref{fig:GB_Indiv_3S_Confusion}, we can see that the emotion pairs of Hate and Disgust as well as Pleasure and Contentment are the most common emotions to be misclassified as one another. We suggest that this is due to these emotions representing similar meanings, thus they would be conveyed using similar features. For instance, Hate and Disgust both tend to consist of lower tones while Pleasure and Contentment have higher tones.
	
	\subsection{Future Work}
	
	One of the major challenges we faced was the limited amount of data  that was collected. We plan on expanding this dataset to a larger variety of singers and other instrumentalists so that we can better understand how the models can generalize to different sounds. Additional future work includes developing a more sophisticated deep-learning based model on the raw audio data for classifying emotion using the expanded dataset we will collect. This will allow the model to make predictions beyond what could be possible using the features we chose in our feature selection. It would open up the potential to achieve much higher accuracy and better model generalization.
	
	\section{Conclusions}\label{sec:conclusions}
	
	Our novel dataset using an expanded emotion taxonomy provides opportunity for the development of a more articulate understanding of emotions. Previous attempts to correlate emotion to audio or music are based on fewer emotions, and often rely  on lyrics or song metadata for classification. Our algorithms demonstrate a high level of accuracy on a 20 category taxonomy for emotions, utilizing only prosodic features. By restricting the type of input data to prosodic features and expanding the number of classified emotions, our models can be used for a wide range of research challenges within the domain of emotional classification. Furthermore, we have demonstrated that our approach is able to generalize to a larger subset of the overall population. Finally, the restriction of our feature vector via additive feature selection demonstrates the ability for prosodic features to achieve a high-level accuracy for emotional classification for  a relatively small number of features. 
	
	
	\bibliography{smc2021bib}
	
\end{document}